\date{}
\documentclass[11pt]{article}
\usepackage{amsmath,amsfonts,latexsym,graphicx,amssymb}
\usepackage{amsfonts}
\usepackage{bm}

\newcommand{\ot}{{\,\otimes\,}}
\newcommand{{\Cd}}{{\mathbb{C}^d}}
\newcommand{\sbsigma}{{\mbox{\scriptsize \boldmath $\sigma$}}}

\newcommand{\bsigma}{{\mbox{ \boldmath $\sigma$}}}

\newcommand{\bmu}{{\mbox{ \boldmath $\mu$}}}
\newcommand{\bnu}{{\mbox{ \boldmath $\nu$}}}

\newcommand{\MD}{\mathfrak{D}}

\newcommand{\sbnu}{{\mbox{\scriptsize \boldmath $\nu$}}}
\def\oper{{\mathchoice{\rm 1\mskip-4mu l}{\rm 1\mskip-4mu l}%
{\rm 1\mskip-4.5mu l}{\rm 1\mskip-5mu l}}}
\def\<{\langle}
\def\>{\rangle}
\newtheorem{theorem}{Theorem}

\numberwithin{equation}{section}

\begin{document}
\title{\textbf{Rotationally invariant multipartite states}} \author{Dariusz
Chru\'sci\'nski and Andrzej Kossakowski \\
Institute of Physics, Nicolaus Copernicus University,\\
Grudzi\c{a}dzka 5/7, 87--100 Toru\'n, Poland}

\maketitle

\begin{abstract}

We construct a  class of multipartite states possessing rotational
SO(3) symmetry -- these are states of $K$ spin-$j_A$ particles and
$K$ spin-$j_B$ particles. The construction of symmetric states
follows our two recent papers devoted to unitary and orthogonal
multipartite symmetry.  We study basic properties of multipartite
SO(3) symmetric states: separability criteria and multi-PPT
conditions.

\end{abstract}

%\pacs{03.65.Ud, 03.67.-a}

\section{Introduction}
\setcounter{equation}{0}

Symmetry plays a prominent role in modern physics. In many cases
it enables one to simplify the analysis of the corresponding
problems and very often it leads to much deeper understanding and
the most elegant mathematical formulation of the corresponding
physical theory. In Quantum Information Theory \cite{QIT} the very
idea of symmetry was first applied by Werner \cite{Werner1} to
construct a highly symmetric family of bipartite $d \ot d$ states
which are invariant under the following local unitary operations
\begin{equation}\label{W}
\rho \ \longrightarrow\  U\ot U \, \rho\, (U \ot U)^\dag\ ,
\end{equation}
where $U$ are unitary operators from $U(d)$ --- the group of
unitary $d \times d$ matrices. Another family of symmetric states
(so called isotropic states \cite{Horodecki}) is governed by the
following invariance rule
\begin{equation}\label{I}
\rho\  \longrightarrow\  U\ot \overline{U} \, \rho \, (U \ot
\overline{U})^\dag\ ,
\end{equation}
where $\overline{U}$ is the complex conjugate of $U$ in some
basis. Other symmetry groups (subgroups of $U(d)$) were  first
considered in \cite{Werner2}.

Let us observe that the problem of symmetric bipartite states may
be formulated in more general setting. Consider the composite
system living in $\mathcal{H}_{\rm total} = \mathcal{H}_A \ot
\mathcal{H}_B$ and let $G$ be  a symmetry group in question. Let
$\MD^{(A)}$ and $\MD^{(B)}$ denote irreducible unitary
representations of $G$ in $\mathcal{H}_A$ and $\mathcal{H}_B$,
respectively. Now, a state $\rho$ of the composite is Werner-like
$\MD^{(A)} \ot \MD^{(B)}$--invariant iff
\begin{equation}\label{}
    [\, \MD^{(A)}(g) \ot \MD^{(B)}(g)\, ,\, \rho \, ] = 0 \ ,
\end{equation}
for all elements $g \in G$. Similarly, $\rho$  is isotropic-like
$\MD^{(A)} \ot \overline{\MD^{(B)}}$--invariant iff
\begin{equation}\label{}
    [\, \MD^{(A)}(g) \ot \overline{\MD^{(B)}(g)}\, ,\, \rho \, ] = 0
    \ .
\end{equation}
It is clear that taking $\mathcal{H}_A = \mathcal{H}_B =
\mathbb{C}^d$ and $\MD^{(A)} = \MD^{(B)} \equiv \MD$ the defining
representation of  $G=U(d)$ one obtains the standard Werner state
\cite{Werner1}. Taking as $G$ a rotational group SO(3) one
constructs a family of rotationally invariant states considered
recently in \cite{Werner2} and in more details in
\cite{R1,R1a,R2,R3} (see also \cite{Diploma}). Rotationally
invariant bipartite states arise from thermal equilibrium states
of low-dimensional spin systems with a rotationally invariant
Hamiltonian by tracing out all degrees of freedom but those two
spins. Entanglement in generic  spin models has recently been
studied in \cite{Spin1,Spin2,Spin3,Spin4}. Rotationally invariant
states were recently applied in quantum optics to describe
multiphoton entangled states produced by parametric
down-conversion \cite{Durkin1} (see also \cite{Durkin2}).

In a present paper we consider a multipartite generalization of
SO(3)-invariant states. Symmetric multipartite states were first
considered in \cite{Werner3} (see also \cite{PhD}) for $G=U(d)$
and $G=O(d)$. An $N$-partite generalization of Werner state in
$H_{\rm total} = (\mathbb{C}^d)^{\ot N}$ is defined by the
following requirement \cite{Werner3}:
\begin{equation}\label{DI}
    [\, U^{\ot N}\, , \, \rho\, ] =0\,
\end{equation}
for all $U\in U(d)$. This definition may be slightly generalized
as follows: an $N$-partite  state $\rho$ living in
$\mathcal{H}_{\rm total} = \mathcal{H}_1 \ot \ldots \ot
\mathcal{H}_N$ is invariant under $\MD^{(1)} \ot \ldots \ot
\MD^{(N)}$, where $\MD^{(k)}$ denotes an irreducible
representation of the symmetry group $G$ in $\mathcal{H}_k$, iff
\begin{equation}\label{}
    [\,\MD^{(1)}(g) \ot \ldots \ot \MD^{(N)}(g)\, , \, \rho\, ] =0\,
\end{equation}
for all $g\in G$.

Recently \cite{I,II} we proposed another family of multipartite
symmetric states. Our construction works for even number of
parties. Consider $K$ copies of $\mathcal{H}_A$ and $K$ copies of
$\mathcal{H}_B$. Let $\MD^{(A)}$ and $\MD^{(B)}$ denote
irreducible unitary representations of $G$ in $\mathcal{H}_A$ and
$\mathcal{H}_B$, respectively. Now, a $2K$-partite state $\rho$ is
$(\MD^{(A)} \ot \ldots \ot \MD^{(A)}) \ot ( \MD^{(B)} \ot \ldots
\ot \MD^{(B)})$--invariant iff
\begin{eqnarray}\label{DII}
    \Big[\, \MD^{(A)}(g_1) \ot \ldots \ot \MD^{(A)}(g_K) \ot  \MD^{(B)}(g_1)
    \ot \ldots \ot \MD^{(B)}(g_K)\, ,\, \rho\, \Big]\,=\,0\ ,
\end{eqnarray}
for all $(g_1,\ldots,g_K)\in G \times \ldots \times G$. Note the
crucial difference between these two definitions (\ref{DI}) and
(\ref{DII}): the first one uses only one element $g$ from $G$
whereas the second one uses $K$ different elements
$g_1,\ldots,g_K$, and hence it is much more restrictive. In
\cite{I} we considered unitary symmetry, i.e. $G=U(d)$ and
$\mathcal{H}_A = \mathcal{H}_B =\mathbb{C}^d$, whereas in
\cite{II} we analyzed orthogonal symmetry with $G=O(d) \subset
U(d)$. It turns out that contrary to the symmetric states
considered in  \cite{Werner3,PhD} the states constructed in
\cite{I,II} give rise to simple separability criteria. In the
present paper we construct multipartite states with rotational
SO(3) symmetry.

The paper is organized as follows: in Section \ref{2-PARTIES} we
recall basic properties of rotationally invariant   bipartite
states. This section summarizes the main results obtained in
\cite{R1,R1a,R2,R3}. In section \ref{MULTI} we construct
multipartite SO(3)-invariant states and study its basic
properties: separability and multi-PPT conditions. More technical
analysis is moved to appendixes. Final conclusions are collected
in the last section.

\section{Rotationally invariant bipartite states}
\label{2-PARTIES} \setcounter{equation}{0}

\subsection{Werner-like states}

 Let us consider  two particles with spins $j_A$ and $j_B\geq j_A$. The
 composed bipartite system lives in
$\mathcal{H}_{AB} = \mathcal{H}_A \ot \mathcal{H}_B$, with
$\mathcal{H}_A = \mathbb{C}^{d_A}$ and $\mathcal{H}_B =
\mathbb{C}^{d_B}$, where $d_A=2j_A +1$ and $d_B = 2j_B +1$. Recall
that the Hilbert space corresponding to spin-$j$ particle is
spanned by $d=2j+1$  eigenstates $|j,m\>$, where
$m=-j,-j+1,\ldots,j$. A bipartite  operator $\rho$ is said to be
Werner-like  rotationally or SO(3)-invariant  iff for any $R \in $
SO(3)
\begin{equation}\label{}
    [\MD^{(j_A)}(R) \ot   \MD^{(j_B)}(R)\, , \rho] = 0\ ,
    % \rho \, \MD^{(j_A)}(R) \ot
    % \MD^{(j_B)}(R)\ ,
\end{equation}
where $\MD^{(j)}(R)$ denotes irreducible unitary representation of
$R$ in $\mathbb{C}^{2j+1}$. As is well known the tensor product of
two irreducible representations $\MD^{(j_A)}(R) \ot
\MD^{(j_B)}(R)$ is no longer irreducible in $\mathbb{C}^{d_A} \ot
\mathbb{C}^{d_B}$. It decomposes into a direct sum of irreducible
representations
\begin{equation}\label{}
\MD^{(j_A)}(R) \ot   \MD^{(j_B)}(R) = \bigoplus_{J = j_B
-j_A}^{j_B + j_A}\, \MD^{(J)}(R)\ ,
\end{equation}
each  appearing with multiplicity 1. The composite space
$\mathcal{H}_{AB}$ is spanned by $d_A\cdot d_B$ vectors $|JM\>$
with $J=j_B -j_A,\ldots,j_B+j_A$ and $M=-J,\ldots,J$, that is
\begin{equation}\label{JM}
    |JM\> = \sum_{m_A,m_B}\, \<j_A,m_A;j_B,m_B|JM\>\, |m_A;m_B\>
\end{equation}
where $\<j_A,m_A;j_B,m_B|JM\>$ denote  Clebsh-Gordan coefficients
\cite{Wigner,Edmonds,Louck}, and
\begin{equation}\label{}
    |m_A;m_B\> = |j_A,m_A\>
    \ot |j_B,m_B\>\ .
\end{equation}
Now, the space of Werner-like SO(3)-invariant operator is spanned
by $2j_A +1$   projectors:
\begin{equation}\label{QJ}
    Q^J = \sum_{M=-J}^J |JM\>\<JM| \ ,
\end{equation}
that is, any SO(3)-invariant operator may be written as follows
\begin{equation}\label{rho-werner-SO3}
    \rho = \sum_J q_J\, \widetilde{Q}^J\ ,
\end{equation}
where $q_J\geq 0$ with $\sum_J q_J =1$, and we use the following
notation $\widetilde{A} = A / \mbox{Tr}\,A$. Note that
$\mbox{Tr}\,Q^J= 2J+1$.

It is evident that an arbitrary bipartite state $\rho$ may be
projected  onto the SO(3)-invariant subspace by the following {\it
twirl} operation:
\begin{equation}\label{cal-D}
    \mathbb{T}(\rho) = \int\, \MD^{(j_A\ot j_B)}(R) \, \rho\, [\MD^{(j_A\ot j_B)}(R)]^\dag\, dR\ ,
\end{equation}
where $d{R}$ is an invariant normalized Haar measure on SO(3), and
we introduce the following slightly more compact notation:
\begin{equation}\label{compact}
\MD^{(j_A\ot j_B)}(R) = \MD^{(j_A)}(R) \ot \MD^{(j_B)}(R)\ .
\end{equation}
 Clearly,
$\mathbb{T}(\rho)$ is of the form (\ref{rho-werner-SO3}) with
fidelities $q_J = \mbox{Tr}(\rho\, Q^J)$.

\subsection{Isotropic-like states}

Now, a bipartite state $\rho$ is isotropic-like SO(3)-invariant
iff
\begin{equation}\label{isotropic-like}
  [  \MD^{(j_A)}(R) \ot   \overline{\MD^{(j_B)}(R)}\, , \rho] = 0\ ,
  % \rho \, \MD^{(j_A)}(R) \ot
  %  \overline{\MD^{(j_B)}(R)}\ ,
\end{equation}
where $\overline{\MD^{(j)}(R)}$ denotes conjugate representation.
Representations  $\MD^{(j)}$ and $\overline{\MD^{(j)}}$ are
equivalent and hence there exists an intertwining unitary operator
$V$ such that $V\MD^{(j)} = \overline{\MD^{(j)}}V$. It turns out
that
\begin{equation}\label{V}
    V|j,m\> = (-1)^{j-m}|j,-m\>\ .
\end{equation}
Let us define a family of projectors
\begin{equation}\label{QVP}
    P^J = (\oper \ot V)Q^J(\oper \ot V^\dag)\ .
\end{equation}
Note, that $P^J$ are $\MD^{(j_A \ot \overline{j_B})}$--invariant,
where in analogy to (\ref{compact}), we introduced
\begin{equation}\label{compact-c}
\MD^{(j_A\ot \overline{j_B})}(R) = \MD^{(j_A)}(R) \ot
\overline{\MD^{(j_B)}(R)}\ .
\end{equation}
 Indeed, one has
\begin{eqnarray}
\MD^{(j_A\ot \overline{j_B})}(R)P^J &=& \MD^{(j_A\ot
\overline{j_B})}(R)(\oper \ot V)Q^J(\oper \ot V^\dag) =
 (\oper \ot V)\MD^{(j_A\ot {j_B})}(R) Q^J(\oper \ot V^\dag) \nonumber \\
&=& (\oper \ot V)Q^J \MD^{(j_A\ot {j_B})}(R) (\oper \ot V^\dag)
= (\oper \ot V)Q^J(\oper \ot V^\dag)  \MD^{(j_A\ot \overline{j_B})}(R) \nonumber \\
&=& P^J \MD^{(j_A\ot \overline{j_B})}(R) \ . \nonumber
\end{eqnarray}
Therefore, any $\MD^{(j_A\ot \overline{j_B})}$--invariant state
has the following form
\begin{equation}\label{rho-iso-SO3}
    \rho = \sum_J p_J\, \widetilde{P}^J\ ,
\end{equation}
where $p_J\geq 0$ with $\sum_J p_J =1$. Again,  an arbitrary
bipartite state $\rho$ may be projected  onto the $\MD^{(j_A\ot
\overline{j_B})}$--invariant subspace by the following twirl
operation:
\begin{equation}\label{cal-D}
    {\mathbb{T}}'(\rho) = \int\, \MD^{(j_A\ot \overline{j_B})}(R) \, \rho\, [\MD^{(j_A\ot \overline{j_B})}(R)]^\dag\, dR\ ,
\end{equation}
where $d{R}$ is an invariant normalized Haar measure on SO(3).
 Clearly,
$\mathbb{T}'(\rho)$ is of the form (\ref{rho-iso-SO3}) with
fidelities $p_J = \mbox{Tr}(\rho\, P_J)$.

\subsection{PPT states}

Note, that both families of SO(3)-invariant states, i.e.
Werner-like states (\ref{rho-werner-SO3}) and isotropic-like
states (\ref{rho-iso-SO3}) are not independent. They are related
by a partial transposition $\oper \ot \tau$, i.e. $\rho$ is
$\MD^{j_A \ot j_B}$--invariant (it belongs to the class
(\ref{rho-werner-SO3})) iff $(\oper \ot \tau)\rho$ is $\MD^{j_A
\ot \overline{j_B}}$--invariant. Equivalently, using twirl
operations $\mathbb{T}$ and $\mathbb{T}'$ one has
\begin{equation}\label{}
    \mathbb{T}' = (\oper \ot \tau) \circ \mathbb{T} \circ (\oper
    \ot \tau) \ .
\end{equation}
for an arbitrary state $\rho$. Now, for any $\MD^{j_A \ot
j_B}$--invariant projector $Q^J$ one has
\begin{equation}\label{}
   (\oper \ot \tau)\widetilde{Q}^J = \sum_{J'}\, {X}_{JJ'} \widetilde{P}_{J'}\ ,
\end{equation}
where the $d_A\times d_A$ matrix $\mathbf{X} =[\mathbf{X}_{JJ'}]$
reads as follows
\begin{equation}\label{XJJ}
{X}_{JJ'} = \mbox{Tr}[ (\oper \ot \tau)\widetilde{Q}^J {P}^{J'} ]\
.
\end{equation}
Note that due to $\sum_J P^J = I_{d_A} \ot I_{d_B}$ one finds
\begin{equation}\label{sum=1}
    \sum_{J'} {X}_{JJ'} = 1\ .
\end{equation}
However the matrix elements ${X}_{JJ'}$ are not necessarily
positive which prevents ${X}$ to be a stochastic matrix.
Interestingly, matrix $X$ satisfies
\begin{equation}\label{XXI}
    {X}^2 = {I}\ ,
\end{equation}
where {I} stands for $d_A \times d_A$ identity matrix which
implies $X^{-1}=X$ (for proof see Appendix~A).

It turns out that using several properties of Clebsch-Gordan
coefficients matrix ${X}_{JJ'}$ may be expressed in terms of so
called 6-$j$ Wigner symbol well known from the quantum theory of
angular momentum \cite{Wigner}.
 Following \cite{R2} we
show in the Appendix~B that ${X}_{JJ'}$ may be expressed as
follows:
\begin{equation}\label{X-6j}
    {X}_{JJ'} =  (-1)^{2j_B}(2J'+1)\,\left\{ \begin{array}{ccc} j_A & j_B &
J \\ j_A & j_B & J' \end{array} \right\}\ ,
\end{equation}
where the curly brackets denote a 6-$j$ Wigner symbol
\cite{Wigner}. Equivalently, using the Racah $W$-coefficients
\begin{equation*}\label{Racah}
    W(j_A,j_B,j_A',j_B';JJ') = (-1)^\alpha\, \left\{ \begin{array}{ccc} j_A & j_B &
J \\ j_A' & j_B' & J' \end{array} \right\}\ ,
\end{equation*}
where $\alpha = {j_A+j_B+j_A'+j_B'}$,  one finds
\begin{equation}\label{XWJJ}
    {X}_{JJ'} = (-1)^{2j_A}(2J'+1)\, W(j_A,j_B,j_A,j_B;JJ')  \ .
\end{equation}
 Therefore, if $\rho$ is given by (\ref{rho-werner-SO3}),
then its partial transposition has the following form:
\begin{equation}\label{}
  (\oper \ot \tau)\rho = \sum_J\, q'_J\, \widetilde{P}^J\ ,
\end{equation}
where
\begin{equation}\label{}
    q'_J = \sum_{J'} q_{J'} \mathbf{X}_{J'J}\ .
\end{equation}
An SO(3)--invariant state (\ref{rho-werner-SO3}) is PPT iff $q'_J
\geq 0$ for all $J=j_B-j_A,\ldots,j_B+j_A$.

Conversely, if $\rho$ is given by (\ref{rho-iso-SO3}), then its
partial transposition has the following form:
\begin{equation}\label{}
  (\oper \ot \tau)\rho = \sum_J\, {p}\, '_J\, \widetilde{Q}^J\ ,
\end{equation}
with
\begin{equation}\label{}
    p\,'_J = \sum_{J'} p_{J'} {X}_{J'J}\ ,
\end{equation}
where we used the fact that ${X}^{-1} = {X}$. An SO(3)--invariant
state (\ref{rho-iso-SO3}) is PPT iff $p\,'_J \geq 0$ for all
$J=j_B-j_A,\ldots,j_B+j_A$.

In Appendix C we show that for $j_B \geq j_A=1/2$ the $2\times 2$
matrix $\bf X$  reads as follows:
\begin{equation}\label{X-2-2}
    {X} = \frac{1}{2j_B+1} \left( \begin{array}{cc} -1&
    2(j_B+1) \\ 2j_B & 1
    \end{array} \right) \ .
\end{equation}
For $j_B\geq j_A=1$ the corresponding $3 \times 3$ matrix $X$ is
given by (Appendix C):
%\begin{widetext}
\begin{equation}\label{X-3-3}
    {X} =  \frac{1}{j_B(j_B+1)(2j_B+1)} \left( \begin{array}{ccc} j_B+1 &
    -(j_B+1)(2j_B+1) & j_B(j_B+1)(2j_B+3) \\ & &  \\- (j_B+1)(2j_B-1) &
    (j_B^2 +j_B-1)(2j_B+1) &
    j_B(2j_B+3) \\ & & \\ j_B(j_B+1)(2j_B-1) &
    j_B(2j_B+1) & j_B
    \end{array} \right) \ .
\end{equation}
%\end{widetext}

\subsection{Separability}

A Werner-like rotationally invariant state $\rho$ is separable iff
there exists a separable state $\sigma$ in $\mathcal{H}_{AB}$ such
that
\begin{equation}\label{}
    \rho = \mathbb{T}(\sigma)\ .
\end{equation}
Moreover, it is clear that pure separable states $\varphi \ot \psi
\in \mathcal{H}_{AB}$ are mapped via twirl into the extremal
separable symmetric states $\mathbb{T}(|\varphi \ot
\psi\>\<\varphi \ot \psi|)$. Note that among invariant projectors
$Q^J$ only one with maximal $J = j_A + j_B$ is separable since
\begin{equation}\label{}
   \widetilde{Q}^{j_A + j_B} = \mathbb{T}(|j_A;j_B\>\<j_A;j_B|)\ .
\end{equation}
 If $J \neq j_A + j_B$ the
corresponding $Q^J$ is not PPT and hence it is not separable. It
is well known \cite{R1,R1a,R2,R3} that for $j_A=1/2$ and arbitrary
$j_B$ rotationally invariant state is separable iff it is PPT,
i.e.
\begin{equation}\label{}
    \rho = q_{j_B-1/2}\widetilde{Q}^{j_B-1/2} + q_{j_B+1/2} \widetilde{Q}^{j_B-1/2}\ ,
\end{equation}
with $q_{j_B-1/2},q_{j_B+1/2}\geq 0$ and $q_{j_B-1/2} +
q_{j_B+1/2}=1$,  is separable iff
\begin{equation}\label{}
q_J' = \sum_{J'=j_B-1/2}^{j_B+1/2}\, q_{J'} {\bf X}_{J'J} \geq 0\
,
\end{equation}
with {\bf X} given in (\ref{X-2-2}). It gives therefore the
following necessary and sufficient condition for separability
\begin{equation}\label{}
    q_{j_B + 1/2} \geq \frac{1}{d_B}  \ .
\end{equation}
Note that PPT states define a convex set -- an interval
$[\mathbf{q},\mathbf{q}']$, with $\mathbf{q}=(0,1)$ and
$\mathbf{q}'=((d_B-1)/d_B,1/d_B)$, where $\mathbf{q} =(q_{j_B -
1/2},q_{j_B + 1/2})$. Clearly, a state corresponding to
$\mathbf{q}$ is separable --- it is $\widetilde{Q}^{j_B + 1/2}$.
To show that a state corresponding to $\mathbf{q}'$ is also
separable let us observe that
\begin{equation*}\label{}
  \mbox{Tr}\left(\sigma\, Q^{j_B+1/2}\right) =  \frac{1}{d_B} \ , \ \ \ \  \mbox{Tr}\left(\sigma\, Q^{j_B-1/2}\right) =
  \frac{d_B-1}{d_B}\ ,
\end{equation*}
where e.g. $\sigma=|-1/2;j_B\>\<-1/2;j_B|$. The same result holds
for $\sigma=|1/2;-j_B\>\<1/2;-j_B|$.

Similarly, an isotropic-like rotationally invariant state in
$\mathbb{C}^2 \ot \mathbb{C}^{d_B}$ is separable iff it is PPT,
that is
\begin{equation}\label{}
    \rho = p_{j_B-1/2}\widetilde{P}^{j_B-1/2} + p_{j_B+1/2} \widetilde{P}^{j_B+1/2}\ ,
\end{equation}
with $p_{j_B-1/2},p_{j_B+1/2}\geq 0$ and $p_{j_B-1/2} +
p_{j_B+1/2}=1$,  is separable iff
\begin{equation}\label{}
    p_{j_B + 1/2} \geq \frac{1}{d_B}  \ .
\end{equation}
Another interesting case is when $j_B \geq j_A=1$. It was shown
\cite{R2,R3} that for integer $j_B$, i.e. odd $d_B=2j_B+1$,
rotationally invariant state is separable iff it is PPT. However,
for half-integer $j_B$ (even $d_B$) there exist bound entangled
states, i.e. PPT but entangled. Now, using (\ref{X-3-3}), for
integer $j_B$ a rotationally invariant state
\begin{equation}\label{}
    \rho = q_{j_B-1}\widetilde{Q}^{j_B-1}  + q_{j_B} \widetilde{Q}^{j_B} +
    q_{j_B+1} \widetilde{Q}^{j_B+1}\ ,
\end{equation}
is separable iff
\begin{eqnarray*}\label{}
    q_{j_B-1}d_B - q_{j_B}(j_B^2-1)  &\leq & 1\ , \\
q_{j_B}(2j_B^2+j_B-1) - q_{j_B-1}(1-2j_B^2+j_B)  &\leq& j_Bd_B \ .
\end{eqnarray*}
The above conditions considerably simplify for $j_B=1$. One
obtains
\begin{equation}\label{}
 q_{0} \leq \frac 13\ , \ \ \ \ \ \ \  q_1 \leq \frac 12\ ,
\end{equation}
which reproduce separability conditions for $O(3)\ot
O(3)$--invariant states (see formula (27) in \cite{II}). Similar
results hold for isotropic-like rotationally invariant states with
$j_B \geq j_A=1$. For $j_B\geq j_A > 1$ the situation is much more
complicated. For some partial results consult \cite{R1,R1a,R2,R3}.

\subsection{Special case: $j_A=j_B$}

Consider now the special case when both particles have the same
spin $j_A=j_B\equiv j$. One has two families of projectors:
\[ Q^0, Q^1, \ldots , Q^{d-1} \ , \]
and
\[ P^0 \equiv P^+_d, P^1, \ldots , P^{d-1} \ , \]
where $d=2j+1$,
\[P^+_d = \frac 1d\, \sum_{m_A,m_B=-j}^j|m_A;m_B\>\<m_A;m_B|\ ,\]
 denotes a
projector onto the maximally entangled state. Using definitions
(\ref{QJ}) and (\ref{QVP}) and properties of the Clebsch-Gordan
coefficients one proves the following
\begin{theorem}
The Schmidt number \cite{SN} of $Q^J$ and $P^J$ is given by
\begin{equation}\label{}
    {\rm SN}(Q^J) = {\rm SN}(P^J) = d-J \ ,
\end{equation}
for $J=0,1,\ldots,d-1$.
\end{theorem}
Note, that in the case of the standard $U\ot U$-invariant Werner
state one has only two projectors: $Q^{d-2}$ and $Q^{d-1}$.
$Q^{d-2}$ has Schmidt number 2 and $Q^{d-1}$ is separable.
Therefore, contrary to the 1-parameter family of Werner states the
$(d-1)$-parameter family of Werner-like SO(3)-invariant states
gives rise to the full {\it spectrum} of entangled states: from
separable one to states with the maximal Schmidt number $d$. In
the case of isotropic $U \ot \overline{U}$-invariant state one has
maximally entangled (i.e. with Schmidt number $d$) $P^0=P^+_d$ and
separable $P^{d-1}$.

The matrix $X_{JJ'}$ given by (\ref{X-6j}) simplifies to
\begin{equation}\label{X-6j}
    {X}_{JJ'} =  (-1)^{d-1}(2J'+1)\,\left\{ \begin{array}{ccc} j & j &
J \\ j & j & J' \end{array} \right\}\ ,
\end{equation}
In particular for $j=1/2$ the formula (\ref{X-2-2}) reconstructs
$X$ matrix for the Werner $U \ot U$-invariant states in
$\mathbb{C}^2 \ot \mathbb{C}^2$ (see formula (15) in \cite{I}):
\begin{equation}\label{}
    {X} = \frac{1}{2} \left( \begin{array}{rc} -1&
    3 \\ 1 & 1
    \end{array} \right) \ .
\end{equation}
For $j=1$ the formula (\ref{X-3-3})  reconstructs $X$ matrix for
the orthogonally $O(3) \ot O(3)$-invariant states in $\mathbb{C}^3
\ot \mathbb{C}^3$ (see formula (30) in \cite{II}):
\begin{equation}\label{}
    {X} = \frac{1}{6} \left( \begin{array}{rrc} 2& - 6 & 10 \\
    - 2 & 3 & 5 \\ 2 & 3 & 1
    \end{array} \right) \ .
\end{equation}

\section{Multipartite SO(3) symmetric states}
\label{MULTI} \setcounter{equation}{0}

\subsection{Werner-like family}

Consider now $2K$--partite system living in $\mathcal{H}_A \ot
\mathcal{H}_B$, where
\begin{equation}\label{}
    \mathcal{H}_A = \mathcal{H}_1 \ot \ldots \ot \mathcal{H}_K \ ,
\end{equation}
and
\begin{equation}\label{}
    \mathcal{H}_B = \mathcal{H}_{K+1} \ot \ldots \ot \mathcal{H}_{2K} \ ,
\end{equation}
with $\mathcal{H}_1 = \ldots = \mathcal{H}_K = \mathbb{C}^{d_A}$
and $\mathcal{H}_{K+1} = \ldots = \mathcal{H}_{2K} =
\mathbb{C}^{d_B}$. Let $\mathbf{R} = (R_1,\ldots,R_K)$ with $R_i
\in SO(3)$ and define
\begin{equation}\label{}
 \MD^{(j_A)}(\mathbf{R}) \ot   \MD^{(j_B)}(\mathbf{R})\, =\,
 \bigotimes_{i=1}^K\, \MD^{(j_A \ot j_B)}(R_i) \ ,
\end{equation}
where for each $i=1,\ldots,K$ a bipartite unitary operator
$\MD^{(j_A \ot j_B)}(R_i)$ acts on $\mathcal{H}_i \ot
\mathcal{H}_{K+i}$. Now, we call a $2K$-partite state a
Werner-like SO(3)-invariant iff
\begin{equation}\label{}
    [ \, \MD^{(j_A)}(\mathbf{R}) \ot   \MD^{(j_B)}(\mathbf{R})\, ,
    \, \rho\, ] = 0\ ,
\end{equation}
for any $\mathbf{R} \in SO(3) \times \ldots \times SO(3)$. To
parameterize the set of $2K$-partite invariant states let us
introduce the following set of projectors:
\begin{equation}\label{}
    \mathbf{Q}^{\bf J}%{\overrightarrow{\!\! J}}
    = Q^{J_1}_{1|K+1} \ot
    \ldots \ot  Q^{J_K}_{K|2K}\ ,
\end{equation}
where $\mathbf{J} = (J_1,\ldots,J_k)$ is a $K$-vector with $J_i =
j_B - j_A, \ldots, j_B + j_A$. It is clear that
\begin{enumerate}
\item $ \mathbf{Q}^{\bf J}$ are SO(3)-invariant,
\item $ \mathbf{Q}^{\bf J} \cdot  \mathbf{Q}^{\bf J'} =
\delta_{\mathbf{J}\mathbf{J}'}\,  \mathbf{Q}^{\bf J}$,
\item $ \sum_{\mathbf{J}}\,  \mathbf{Q}^{\bf J} = (I_{d_A} \ot
I_{d_B})^{\ot K}$.
\end{enumerate}
Therefore, an arbitrary $2K$-partite SO(3)-invariant state has the
following form
\begin{equation}\label{}
    \rho = \sum_{\mathbf{J}}\, q_{\mathbf{J}}  \widetilde{\mathbf{Q}}^{\bf J}\
    ,
\end{equation}
with $q_{\mathbf{J}}\geq 0$ and $\sum_{\mathbf{J}}\,
q_{\mathbf{J}} =1$. Hence, the set of rotationally invariant
states defines $(d_A^K -1)$-dimensional simplex.

\subsection{Isotropic-like family}

It is clear that we may use the same scheme to define $2k$-partite
isotropic-like states. For any $\mathbf{R} = (R_1,\ldots,R_K)$
with $R_i \in SO(3)$ one defines
\begin{equation}\label{}
 \MD^{(j_A)}(\mathbf{R}) \ot   \overline{\MD^{(j_B)}(\mathbf{R})}\, =\,
 \bigotimes_{i=1}^K\, \MD^{(j_A \ot \overline{j_B})}(R_i) \ ,
\end{equation}
where for each $i=1,\ldots,K$ a bipartite unitary operator
$\MD^{(j_A \ot \overline{j_B})}(R_i)$ acts on $\mathcal{H}_i \ot
\mathcal{H}_{K+i}$. Now, we call a $2K$-partite state an
isotropic-like SO(3)-invariant iff
\begin{equation}\label{}
    [ \, \MD^{(j_A)}(\mathbf{R}) \ot   \overline{\MD^{(j_B)}(\mathbf{R})}\, ,
    \, \rho\, ] = 0\ ,
\end{equation}
for any $\mathbf{R} \in SO(3) \times \ldots \times SO(3)$. To
parameterize the set of $2K$-partite invariant states let us
introduce the following set of projectors:
\begin{equation}\label{}
    \mathbf{P}^{\bf J}%{\overrightarrow{\!\! J}}
    = P^{J_1}_{1|K+1} \ot
    \ldots \ot  P^{J_K}_{K|2K}\ ,
\end{equation}
where $\mathbf{J} = (J_1,\ldots,J_k)$ is a $K$-vector with $J_i =
j_B - j_A, \ldots, j_B + j_A$. It is clear that
\begin{enumerate}
\item $ \mathbf{P}^{\bf J}$ are SO(3)-invariant,
\item $ \mathbf{P}^{\bf J} \cdot  \mathbf{P}^{\bf J'} =
\delta_{\mathbf{J}\mathbf{J}'}\,  \mathbf{P}^{\bf J}$,
\item $ \sum_{\mathbf{J}}\,  \mathbf{P}^{\bf J} = (I_{d_A} \ot
I_{d_B})^{\ot K}$.
\end{enumerate}
Therefore, an arbitrary $2K$-partite SO(3)-invariant state has the
following form
\begin{equation}\label{iso-multi}
    \rho = \sum_{\mathbf{J}}\, p_{\mathbf{J}}\,  \widetilde{\mathbf{P}}^{\bf J}\
    ,
\end{equation}
with $p_{\mathbf{J}}\geq 0$ and $\sum_{\mathbf{J}}\,
p_{\mathbf{J}} =1$. The set of rotationally invariant states
defines $(d_A^K -1)$-dimensional simplex.

\subsection{$\bsigma$-PPT states}

Now, following \cite{I} let us introduce the family of partial
transpositions parameterized by a binary $K$-vector
$\bsigma=(\sigma_1,\ldots,\sigma_K)$:
\begin{equation}\label{}
    \tau_\sbsigma = \oper^{\ot K} \ot \tau^{\sigma_1} \ot \ldots
    \ot \tau^{\sigma_K} \ ,
\end{equation}
where $\tau^\alpha = \oper$ for $\alpha=0$ and $\tau^\alpha =\tau$
for $\alpha=1$. A $2K$-partite state $\rho$ is $\bsigma$-PPT iff
$\tau_\sbsigma \rho \geq 0$. In terms of coefficients
$q_{\mathbf{J}}$ the property of $\bsigma$-PPT leads to the
following conditions
\begin{equation}\label{}
    \sum_{\mathbf{J}}\, q_\mathbf{J}\,
    \mathbf{X}_{\mathbf{J}\mathbf{J}'}^{\sbsigma} \geq 0 \ ,
\end{equation}
for all $\mathbf{J}'$. The $d_A^K \times d_A^K$ matrix
$\mathbf{X}_{\mathbf{J}\mathbf{J}'}^{\sbsigma}$ is given by
\begin{equation}\label{}
\mathbf{X}_{\mathbf{J}\mathbf{J}'}^{\sbsigma} \, =\,
\mbox{Tr}\Big[ (\tau_\sbsigma\widetilde{\bf Q}^{\bf J})\cdot {\bf
P}_{{\bf J}'} \Big]\ .
\end{equation}
 Let us observe that
\begin{equation}\label{}
\mathbf{X}^{\sbsigma} \, =\, {X}^{\sigma_1} \ot \ldots \ot
{X}^{\sigma_K}\ ,
\end{equation}
where ${X}$ is defined by in (\ref{XWJJ}). In component notation
one finds
\begin{equation}\label{}
\mathbf{X}_{\mathbf{J}\mathbf{J}'}^{\sbsigma} \, =\,
{X}^{\sigma_1}_{J_1J_1'}  \ldots {X}^{\sigma_K}_{J_KJ_K'}\ ,
\end{equation}
 Again, in analogy to (\ref{sum=1}) and
(\ref{XXI}) one has
\begin{equation}\label{}
\sum_{J'} \mathbf{X}^\sbsigma_{\mathbf{J}\mathbf{J}'} = 1\ ,
\end{equation}
and $\, \mathbf{X}^\sbsigma \cdot \mathbf{X}^\sbsigma \, =\, {
I}^{\ot K}$ for any $\bsigma$. Therefore
\begin{equation}\label{X-1X}
(\mathbf{X}^\sbsigma)^{-1} = \mathbf{X}^\sbsigma\ .
\end{equation}

In the same way one defines a $\bsigma$-PPT subset of
isotropic-like $2K$-partite symmetric states. A state $\rho$ from
the family (\ref{iso-multi}) is $\bsigma$-PPT iff
$\tau_\sbsigma\rho\geq 0$, that is
\begin{equation}\label{}
    \sum_{\mathbf{J}}\, p_\mathbf{J}\,
    \mathbf{X}^{\sbsigma}_{\mathbf{J}\mathbf{J}'} \geq 0 \ ,
\end{equation}
for all $\mathbf{J}'$.

\subsection{$\bsigma$-invariance}

Note, that each binary vector $\bsigma$ gives rise to the new
$2K$-partite family of symmetric states. We call a state $\rho$
$\bsigma$-invariant iff $\tau_\sbsigma\rho$ is Werner-like
invariant. To parameterize this family let us introduce the
following set of bipartite operators:
\begin{equation}\label{}
\Pi^{J}_{(\sigma)} = \left\{ \begin{array}{ll} {Q}^{J} \ , & \ \
\sigma=0
\\ P^{J} \ ,& \ \ \sigma=1
\end{array} \right. \ \ .
\end{equation}
This operators may be used to construct a set of $2K$-partite
projectors
\begin{equation}\label{}
    \mathbf{\Pi}^{\bf J}_{(\sbsigma)}
    = \Pi^{J_1}_{(\sigma_1)1|K+1} \ot
    \ldots \ot  \Pi^{J_K}_{(\sigma_K)K|2K}\ ,
\end{equation}
satisfying
\begin{enumerate}
\item $ \mathbf{\Pi}^{\bf J}_{(\sbsigma)}$ are $\bsigma$-invariant,
\item $ \mathbf{\Pi}^{\bf J}{(\sbsigma)} \cdot  \mathbf{\Pi}^{\bf J'}_{(\sbsigma)}=
\delta_{\mathbf{J}\mathbf{J}'}\,  \mathbf{\Pi}^{\bf
J}_{(\sbsigma)}$,
\item $ \sum_{\mathbf{J}}\,  \mathbf{\Pi}^{\bf J}_{(\sbsigma)} = (I_{d_A} \ot
I_{d_B})^{\ot K}$.
\end{enumerate}
Therefore, an arbitrary $2K$-partite $\bsigma$-invariant state has
the following form
\begin{equation}\label{iso-multi}
    \rho = \sum_{\mathbf{J}}\, \pi_{\mathbf{J}}\,  \widetilde{\mathbf{\Pi}}^{\bf J}_{(\sbsigma)}\
    ,
\end{equation}
with $\pi_{\mathbf{J}}\geq 0$ and $\sum_{\mathbf{J}}\,
\pi_{\mathbf{J}} =1$. Clearly, the set of $\bsigma$-invariant
states defines $(d_A^K -1)$-dimensional simplex. Let us note that
for any two binary vectors $\bmu$ and $\bnu$ if $\rho$ is
$\bmu$-invariant then $\tau_\sbnu\rho$ is $(\bmu
\oplus\bnu)$-invariant, where $\bmu \oplus\bnu$ denotes addition
mod 2.

\subsection{Separability}

A $2K$-partite Werner-like rotationally invariant state $\rho$ is
separable iff there exists a separable state $\sigma$ in
$\mathcal{H}_{\rm total}$ such that
\begin{equation}\label{}
    \rho = \mathbb{T}_K(\sigma)\ ,
\end{equation}
where $\mathbb{T}_K$ denotes $2K$-partite twirl operation:
\begin{eqnarray*}\label{}
\mathbb{T}_K(\omega) \, = \, \int   \Big[\,
\MD^{(j_A)}(\mathbf{R}) \ot \MD^{(j_B)}(\mathbf{R})\, \Big] \,
    \omega \, \Big[\, \MD^{(j_A)}(\mathbf{R}) \ot
    \MD^{(j_B)}(\mathbf{R})\, \Big]^\dag\, dR_1 \ldots dR_K\ .
\end{eqnarray*}
Moreover, it is clear that pure separable states $\varphi_1 \ot
\ldots \ot \varphi_K \ot \psi_1 \ldots \ot \psi_K \in
\mathcal{H}_{\rm total}$ are mapped via twirl $\mathbb{T}_K$ into
the extremal separable symmetric states. Again only one invariant
projector $\mathbf{Q}^\mathbf{J}$ is separable
--- that corresponding to $\mathbf{J} = (j_A + j_B, \ldots , j_A +
j_B)$. It is given by the twirl of $|j_A\> \ot \ldots \ot |j_A\>
\ot |j_B\> \ot \ldots \ot |j_B\>$.

Using techniques applied in \cite{I,II} one easily shows the
following
\begin{theorem}
If $j_B\geq j_A =1/2$ or  $j_B\geq j_A =1$ with integer $j_B$,
then an arbitrary $\bmu$-invariant state $\rho$ is fully separable
iff it is $\bnu$-PPT for all binary $K$-vectors $\bnu$. Moreover
$\rho$ is $(1\ldots K|K+1\ldots 2K)$ biseparable iff it is
(1\ldots 1)-PPT.
\end{theorem}

In particular for $j_A=j_B=1/2$  the above theorem reconstructs
separability conditions for $\bmu$-invariant states with unitary
symmetry $U(2)$, see \cite{I}, whereas for $j_A=j_B=1$ one
reconstructs separability conditions for $O(3)$-invariant states,
see \cite{II}.

\subsection{Reductions}

It is evident that reducing the $2K$ partite $\bsigma$--invariant
state with respect to $\mathcal{H}_i \ot \mathcal{H}_{i+K}$ pair
one obtains $2(K-1)$--partite $\bsigma_{(i)}$--invariant state
with
\begin{equation}
\bsigma_{(i)} =
(\sigma_1,\ldots,\check{\sigma}_i,\ldots,\sigma_K)\ ,
\end{equation}
where $\check{\sigma}_i$ denotes the omitting of $\sigma_i$. The
reduced state lives in
\begin{equation}\label{}
    \mathcal{H}_1 \ot \ldots \check{\mathcal{H}}_i \ot \ldots \ot
    \check{\mathcal{H}}_{i+K} \ot \ldots \ot \mathcal{H}_{2K}\ .
\end{equation}
The corresponding fidelities of the reduced symmetric state are
given by
\begin{equation}\label{}
    \pi_{(J_1\ldots J_{K})} =
    \sum_{j=j_B-j_A}^{j_B+j_A}\,
    \pi_{(J_1\ldots J_{i-1}jJ_{i+1}\ldots J_{K})}\
    .
\end{equation}
Note, that reduction with respect to a `mixed' pair, say
$\mathcal{H}_i \ot \mathcal{H}_{j+K}$ with $j\neq i$ ($i,j\leq
K$), is equivalent to two `natural' reductions with respect to
$\mathcal{H}_i \ot \mathcal{H}_{i+K}$ and $\mathcal{H}_j \ot
\mathcal{H}_{j+K}$ and hence it gives rise to $2(K-2)$--partite
invariant state. This procedure establishes a natural hierarchy of
multipartite invariant states.

\section{Conlusions}
\setcounter{equation}{0}

 We have introduced a new family of
multipartite rotationally symmetric states for $2K$ particles: $K$
spin-$j_A$ and $K$ spin-$j_B$ particles ($j_B \geq j_A$). Within
this class we have formulated separability conditions for $j_B
\geq j_A=1/2$ and $j_B \geq j_A=1$ with integer $j_B$. It turned
out that full $2K$-separability is equivalent to multi
$\bsigma$-PPT conditions with $\bsigma$ being a binary $K$-vector.

 Recently, a detailed analysis of
multipartite symmetric states and their application in  quantum
information theory was performed by Eggeling in his  PhD thesis
\cite{PhD}. This construction may be applied for SO(3) symmetry as
follows. Consider $N$ spin-$j$ particles. An $N$-partite state
$\rho$ is rotationally invariant iff
\begin{equation}\label{WE}
    [\,\MD^{(j)}(R) \ot \ldots \ot \MD^{(j)}(R)\, , \, \rho\, ] =0\,
\end{equation}
for all $R \in SO(3)$. It is clear that the detailed
parametrization of this class is highly nontrivial: it corresponds
to addition of $N$ angular momenta and, as is well known even the
case $N=3$ gives rise to considerable complications (see e.g.
\cite{Wigner,Edmonds,Louck}). If $N=2K$ our class defines only a
commutative  subclass within Eggeling's class.

Note, that our construction may be slightly generalized. Instead
of $K$ spin-$j_A$ and $K$ spin-$j_B$ particles we may consider
$2K$ particles with arbitrary spins: \[ (j_{A_1},j_{B_1}), \
(j_{A_2},j_{B_2}), \ \ldots ,\ (j_{A_K},j_{B_K})\ . \] Now, a
$2K$-partite state $\rho$ is SO(3)-invariant iff
\begin{equation}\label{}
    \left[ \, \bigotimes_{i=1}^K \MD^{(j_{A_i})}(R_i) \ot  \bigotimes_{i=1}^K \MD^{(j_{B_i})}({R_i})\, ,
    \, \rho\, \right] = 0\ ,
\end{equation}
for all $R_1,\ldots,R_K \in SO(3)$. It is clear that such general
situation does not apply for Eggeling's construction where all
particles carry the same spins.

It is hoped that the multipartite state constructed in this paper
may serve as a  laboratory for testing various concepts from
quantum information theory and they may shed new light on the more
general investigation of multipartite entanglement. Note, that
using duality between bipartite quantum  states and quantum
channels \cite{AJ} one may consider rotationally invariant quantum
channels transforming a state of spin-$j_B$ particle into a state
of spin-$j_A$ one. Relaxing positivity condition upon $\rho$ the
above duality gives rise to rotationally invariant positive maps
which may be used to detect quantum bipartite entanglement. In the
multipartite case the situation is different. Now a crucial role
is played by maps which are positive but only on separable states.
Note that a tensor product of two positive maps is no longer
positive but clearly it is positive on separable states.
Therefore, our  construction of multipartite symmetric states may
be dually used to produce invariant classes of multi-linear maps
which may serve as a useful tool in detecting multi-partite
entanglement.

\section*{Appendix A}
\def\theequation{A.\arabic{equation}}
\setcounter{equation}{0}

Using properties of the operator $V$ defined in (\ref{V}) one
shows that
\begin{equation}\label{A}
    \mbox{Tr}\left[ (\oper \ot \tau)Q^J P^{J'}\right] =  \mbox{Tr}\left[ (\oper \ot \tau)Q^{J'}
    P^{J}\right]\ ,
\end{equation}
that is,
\begin{equation}\label{}
    (2J+1) X_{JJ'} = (2J'+1)X_{J'J}\ .
\end{equation}
 Now, following (\ref{XJJ}) one has
\begin{eqnarray}\label{XJJ+}
{X}^{-1}_{JJ'}& =& \mbox{Tr}\left[ (\oper \ot \tau)\widetilde{P}^J
{Q}^{J'} \right] \nonumber \\ &=& \frac{1}{2J+1}\, \mbox{Tr}\left[
(\oper \ot \tau){P}^J {Q}^{J'} \right] \ ,
\end{eqnarray}
and using (\ref{A}) one shows that ${X}^{-1}_{JJ'} = {X}_{JJ'}$,
that is, $X^2=I$.

\section*{Appendix B}
\def\theequation{B.\arabic{equation}}
\setcounter{equation}{0}

Using (\ref{QJ}) and (\ref{QVP}) one finds for the matrix $\bf X$:
%\begin{widetext}
\begin{eqnarray}\label{}
  \mathbf{X}_{JJ'} = \frac{1}{2J+1} \sum_{M,M'}\mbox{Tr}\Big[ (\oper \ot
    \tau)|JM\>\<JM| (\oper \ot V)|J'M'\>\<J'M'|(\oper \ot V^\dag)
    \Big]\ .
\end{eqnarray}
%\end{widetext}
Therefore, taking into account (\ref{JM}) and the following
 relation between Clebsch-Gordan coefficients and 3-$j$ Wigner
symbols:
\begin{equation}\label{}
    \< j_1,j_2;m_1,m_2|JM\> = (-1)^{j_1-j_2 +M} \sqrt{2J+1} \left( \begin{array}{ccc} j_1 & j_2
    & J
\\ m_1 & m_2 & -M \end{array} \right) \ ,
\end{equation}
one obtains
%\begin{widetext}
\begin{eqnarray*}\label{}
 && \mbox{Tr}\Big[ (\oper \ot
    \tau)|JM\>\<JM| (\oper \ot V)|J'M'\>\<J'M'|(\oper \ot V^\dag)
    \Big] \nonumber \\ && =\, (2J+1)(2J'+1) \sum_{m_A,m_B}\sum_{l_A,l_B}
    \sum_{m_A',m_B'}\sum_{l_A',l_B'} (-1)^{2(M+M')}\, \delta_{m_A,l_A'}\delta_{l_A, m_A'}
\delta_{m_B,-m_B'}\delta_{l_B,-l_B'}  \nonumber \\
 && \times\,
\left( \begin{array}{ccc} j_A & j_B & J
\\ m_A & m_B & -M \end{array} \right)
\left( \begin{array}{ccc} j_A & j_B & J
\\ l_A & l_B & -M \end{array} \right)
\left( \begin{array}{ccc} j_A & j_B & J'
\\ m_A' & m_B' & -M' \end{array} \right)
\left( \begin{array}{ccc} j_A & j_B & J'
\\ l_A' & l_B' & -M' \end{array} \right) \ .
\end{eqnarray*}
Finally, using the symmetry of 3-$j$ symbols
\begin{equation}\label{}
\left( \begin{array}{ccc} l_1 & l_2 &l_3
\\ m_1 & m_2 & m_3 \end{array} \right) = (-1)^{l_1 + l_2 + l_3}\left( \begin{array}{ccc} l_1 & l_2 &l_3
\\ -m_1 & -m_2 & -m_3 \end{array} \right)\ ,
\end{equation}
and the following relation between 3-$j$ and 6-$j$ symbols
\begin{eqnarray}
&& (-1)^{l_1' + l_2' + l_3'} \left\{ \begin{array}{ccc} l_1 & l_2
& l_3 \\ l_1' & l_2' & l_3' \end{array} \right\} \, =\,
\sum_{m_1,m_1'} \sum_{m_2,m_2'} \sum_{m_3,m_3'} (-1)^{m_1' + m_2'
+ m_3'} \nonumber \\ & \times& \!\!\left(
\begin{array}{ccc} l_1 & l_2 &l_3
\\ m_1 & m_2 & m_3 \end{array} \right)
\left( \begin{array}{ccc} l_1 & l_2' &l_3'
\\ -m_1 & m_2' & -m_3' \end{array} \right) %\nonumber \\
 %\hspace{3cm}
 \left( \begin{array}{ccc} l_1' & l_2 &l_3'
\\ -m_1' & -m_2 & m_3' \end{array} \right)
\left( \begin{array}{ccc} l_1' & l_2' &l_3
\\ m_1' & -m_2' & -m_3 \end{array} \right) \ ,
\end{eqnarray}
%\end{widetext}
one proves (\ref{X-6j}).

\section*{Appendix C}
\def\theequation{C.\arabic{equation}}
\setcounter{equation}{0}

The 6-$j$-symbols are invariant under permutation of their
columns, e.g.
\begin{equation}\label{S1}
\left\{ \begin{array}{ccc} j_1 & j_2 & j_3 \\ J_1 & J_2 & J_3
\end{array} \right\}  = \left\{ \begin{array}{ccc} j_2 & j_1 &
j_3 \\ J_2 & J_1 & J_3 \end{array} \right\}  \ ,
\end{equation}
and under exchange of two corresponding elements between rows,
e.g.
\begin{equation}\label{S2}
\left\{ \begin{array}{ccc} j_1 & j_2 & j_3 \\ J_1 & J_2 & J_3
\end{array} \right\}  = \left\{ \begin{array}{ccc} J_1 & J_2 & j_3 \\ j_1 & j_2 & J_3
\end{array} \right\} \ .
\end{equation}
 Now, to find the $\bf X$ matrix for
$j_A=1$ we shall use the following formulae \cite{Wigner}:
%\begin{widetext}
\begin{equation}\label{}
\left\{ \begin{array}{ccc} j_1 - \frac 12 & \frac 12 & j_1 \\
j_2 & j & j_2 - \frac 12
\end{array} \right\}  = (-1)^J \left[
\frac{(J+1)(J-2j)}{2j_1(2j_1+1)2j_2(2j_2+1)}\right]^{1/2}\ ,
\end{equation}
\begin{equation}\label{}
\left\{ \begin{array}{ccc} j_1 - \frac 12 & \frac 12 & j_1 \\
j_2 - \frac 12 & j & j_2
\end{array} \right\}  = (-1)^{J - 1/2} \left[
\frac{(J-2j_1+\frac 12)(J-2j_2 + \frac
12)}{2j_1(2j_1+1)2j_2(2j_2+1)}\right]^{1/2}\ ,
\end{equation}
%\end{widetext}
with $J=j_1+j_2+j$. Using these formulae together with symmetry
relations (\ref{S1})--(\ref{S2}) one obtains (\ref{X-2-2}).

 Now, to find the $\bf X$ matrix for
$j_A=1$ we shall use the following formulae \cite{Wigner}:
%\begin{widetext}
\begin{eqnarray}\label{}
 \left\{ \begin{array}{ccc} j_1 - 1 &  1 & j_1 \\
j_2  & j & j_2 -1
\end{array} \right\} & =&  (-1)^{J} \left[
\frac{J(J+1)(J-2j
-1)(J-2j)}{(2j_1-1)2j_1(2j_1+1)(2j_2-1)2j_2(2j_2+1)}\right]^{1/2}\
, \\
%\end{equation}
%%
%\begin{eqnarray}\label{}
\left\{ \begin{array}{ccc} j_1 - 1 &  1 & j_1 \\
j_2-1  & j & j_2
\end{array} \right\}  &=& (-1)^{J-1} \left[
\frac{(J-2j_1)(J-2j+1)(J-2j_2)(J-2j_2+1)}
{(2j_1-1)2j_1(2j_1+1)(2j_2-1)2j_2(2j_2+1)}\right]^{1/2}\ , \\
%\end{equation}
%%
%\begin{equation}\label{}
\left\{ \begin{array}{ccc} j_1  &  1 & j_1 \\
j_2-1  & j & j_2
\end{array} \right\}  &=& (-1)^{J} \left[
\frac{2(J+1)(J-2j)(J-2j_1)(J-2j_2+1)}
{2j_1(2j_1+1)(2j_1+2)(2j_2-1)2j_2(2j_2+1)}\right]^{1/2}\ , \\
%\end{equation}
%%
%\begin{equation}\label{}
\left\{ \begin{array}{ccc} j_1  &  1 & j_1 \\
j_2  & j & j_2
\end{array} \right\}  &=& (-1)^{J}
\frac{j(j+1) -j_1(j_1+1)- j_2(j_2+1)} {[
j_1(2j_1+1)(2j_1+2)j_2(2j_2+1)(2j_2+2)]^{1/2}}\ ,
\end{eqnarray}
%\end{widetext}
together with symmetry relations (\ref{S1})--(\ref{S2}). Simple
calculations give (\ref{X-3-3}).

\section*{Acknowledgments}  After this work was completed we have
learnt from Gabriel Durkin that  rotationally invariant states
were studied recently in his PhD thesis \cite{Durkin2}. We thank
him for sending us his thesis.  This work was partially supported
by the Polish State Committee for Scientific Research Grant {\em
Informatyka i in\.zynieria kwantowa} No PBZ-Min-008/P03/03.


\begin{thebibliography}{1} \bibliographystyle{plain}

\bibitem{QIT} M. A. Nielsen and I. L. Chuang, {\it Quantum computation
and quantum information}, Cambridge University Press, Cambridge,
2000.

\bibitem{Werner1} R.F. Werner, Phys. Rev. A {\bf 40}, 4277 (1989).

\bibitem{Horodecki} M. Horodecki and P. Horodecki, Phys. Rev. {\bf A} 59, 4206
(1999).


\bibitem{Werner2} K.G.H. Vollbrecht and R.F. Werner, Phys. Rev. A {\bf 64}, 062307 (2001).


\bibitem{R1} J. Schliemann, Phys. Rev. A {\bf 68}, 012309 (2003).

\bibitem{R1a} J. Schliemann,
Phys. Rev. A {\bf 72}, 012307 (2005).

\bibitem{R2} H.-P. Breuer,  Phys. Rev. A {\bf 71}, 062330 (2005).

\bibitem{R3} H.-P. Breuer, J. Phys. A: Math. Gen. {\bf 38}, 9019
(2005).

\bibitem{Diploma} SU(2)-invariant states were also studied by B. Hendriks and R.
F. Werner [B. Hendriks, Diploma thesis, University of
Braunschweig, Germany, 2002].

%------------------------SPIN-----------------------------

\bibitem{Spin1} A. Osterloh, L. Amico, G. Falci, and R. Fazio, Nature {\bf 416}, 608
(2002).

\bibitem{Spin2} T. J. Osborne and M. A. Nielsen, Phys. Rev. A {\bf 66}, 032110
(2002).

\bibitem{Spin3} F. Verstraete, M. Popp, and J. I. Cirac, Phys. Rev. Lett. {\bf 92},
027901 (2004).

\bibitem{Spin4} W. D\"{u}r, L. Hartmann, M. Hein, M. Lewenstein and H. J. Briegel,
Phys. Rev. Lett. {\bf 94}, 097203 (2005).

\bibitem{Durkin1} G.A. Durkin, C. Simon, J. Eisert and  D. Bouwmeester,  Phys.
Rev. A {\bf 70}, 062305 (2004).

\bibitem{Durkin2} G.A. Durkin, {\it Light and Spin Entanglement},  PhD thesis,
Oxford, 2004.


%----------------MULTI------------------------------------

\bibitem{Werner3} T. Eggeling and R.F. Werner, Phys. Rev. A {\bf 63}, 042111 (2001).

\bibitem{PhD}  T. Eggeling, {\it On Multipartite Symmetric
States in Quantum Information Theory}, PhD thesis, Braunschweig,
2003 (available on line at
www.biblio.tu-bs.de/ediss/data/20030602b/20030602b.pdf).


\bibitem{I} D. Chru\'sci\'nski and A. Kossakowski, Phys. Rev. A
{\bf 73}, 062314 (2006).



\bibitem{II} D. Chru\'sci\'nski and A. Kossakowski, Phys. Rev. A
{\bf 73}, 062315 (2006).

%------------------------ANGULAR-------------------------

\bibitem{Wigner} E.P. Wigner, {\it Group Theory and its
Application to the Quantum Mechanics of Atomic Spectra}, (Academic
Press, New York, 1959).

\bibitem{Edmonds} A.R. Edmonds, {\it Angular Momentum in Quantum
Mechanics}, (Princeton University Press, Princeton, 1957).

\bibitem{Louck} L.C. Biedenharn and J.D. Louck, {\it Angular
Momentum in Quantum Physics: Theory and Applications},
(Addison-Wesley Publishing Company Reading, Massachusetts, 1981).

%-------------------------------------------------------------

\bibitem{SN} B. Terhal and P. Horodecki,
Phys. Rev. A {\bf 61}, 040301 (2000).

\bibitem{AJ} A. Jamio{\l}kowski, Rep. Math. Phys. {\bf 3}, 275
(1972).



\end{thebibliography}
\end{document}